\documentclass{article}

\usepackage{spconf,amsmath}
\usepackage{hhline}
\usepackage{scalerel} 

\usepackage[utf8]{inputenc} 
\usepackage[T1]{fontenc}    
\usepackage{hyperref}       
\usepackage{url}            
\usepackage{booktabs}       
\usepackage{amsfonts}       
\usepackage{nicefrac}       
\usepackage{microtype}      

\usepackage{subcaption}
\usepackage{multicol}
\usepackage{algorithm}
\usepackage{epsfig}
\usepackage{graphicx}
\usepackage{amsmath}
\usepackage{amssymb}
\usepackage{booktabs} 

\title{Generalized Octave Convolutions for\\Learned Multi-Frequency Image Compression}

\name
{Mohammad Akbari$^*$, Jie Liang$^*$, Jingning Han$^\dagger$, Chengjie Tu$^\ddagger$}
\address{akbari@sfu.ca, jiel@sfu.ca, jingning@google.com, chengjietu@tencent.com\\Simon Fraser University, Canada$^*$, Google Inc.$^\dagger$, Tencent Technologies$^\ddagger$}

\begin{document}\sloppy

\def\x{{\mathbf x}}
\def\L{{\cal L}}

%

\maketitle

\begin{abstract}
Learned image compression has recently shown the potential to outperform the standard codecs. State-of-the-art rate-distortion (R-D) performance has been achieved by context-adaptive entropy coding approaches in which hyperprior and autoregressive models are jointly utilized to effectively capture the spatial dependencies in the latent representations. However, the latents are feature maps of the same spatial resolution in previous works, which contain some redundancies that affect the R-D performance. In this paper, we propose the first learned multi-frequency image compression and entropy coding approach that is based on the recently developed octave convolutions to factorize the latents into high and low frequency (resolution) components, where the low frequency is represented by a lower resolution. Therefore, its spatial redundancy is reduced, which improves the R-D performance. Novel generalized octave convolution and octave transposed-convolution architectures with internal activation layers are also proposed to preserve more spatial structure of the information. Experimental results show that the proposed scheme not only outperforms all existing learned methods as well as standard codecs such as the next-generation video coding standard VVC (4:2:0) on the Kodak dataset in both PSNR and MS-SSIM. We also show that the proposed generalized octave convolution can improve the performance of other auto-encoder-based computer vision tasks such as semantic segmentation and image denoising.
\end{abstract}
\begin{keywords}
generalized octave convolutions, multi-frequency autoencoder, multi-frequency image coding, learned image compression, learned entropy model
\end{keywords}

\section{Introduction}
\label{sec:introduction}

Deep learning-based image compression \cite{rippel2017real, agustsson2017soft,akbari2019dsslic,balle2016,johnston2017improved,li2019learning,minnen2018joint,lee2018context,theis2017lossy,toderici2015variable,toderici2017full,li2020deep} has shown the potential to outperform standard codecs such as JPEG2000 and H.265/HEVC-based BPG image codec \cite{bellard2017bpg}.
Learned image compression was first used in \cite{toderici2015variable} to compress thumbnail images using long short-term memory (LSTM)-based recurrent neural networks (RNNs) in which better SSIM results than JPEG and WebP were reported. This approach was generalized in \cite{johnston2017improved}, which utilized spatially adaptive bit allocation to further improve the performance. 

In \cite{balle2016}, a scheme based on generalized divisive normalization (GDN) and inverse GDN (IGDN) were proposed, which outperformed JPEG2000 in both PSNR and SSIM. A compressive auto-encoder framework with residual connection as in ResNet was proposed in \cite{theis2017lossy}, where the quantization was replaced by a smooth approximation, and a scaling approach was used to get different rates. In \cite{agustsson2017soft}, a soft-to-hard vector quantization approach was introduced, and a unified framework was developed for image compression as well as deep learning model compression. In order to take the spatial variation of image content into account, a content-weighted framework was introduced in \cite{li2019learning}, where an importance map for locally adaptive bit rate allocation was employed. A learned channel-wise quantization along with arithmetic coding was also used to reduce the quantization error. 

There have also been some efforts in taking advantage of other computer vision tasks in image compression frameworks. For example, in \cite{akbari2019dsslic}, a deep semantic segmentation-based layered image compression (DSSLIC) was proposed, by taking advantage of the Generative Adversarial Network (GAN) and BPG-based residual coding. It outperformed the BPG codec (in RGB444 format) in both PSNR and MS-SSIM \cite{wang2003multiscale} across a large range of bit rates.

Since most learned image compression methods need to train multiple networks for multiple bit rates, variable-rate approaches have also been proposed in which a single neural network model is trained to operate at multiple bit rates. This approach was first introduced by \cite{toderici2015variable}, which was then generalized for full-resolution images using deep learning-based entropy coding in \cite{toderici2017full}. A CNN-based multi-scale decomposition transform was optimized for all scales in \cite{cai2018efficient}, which achieved better performance than BPG in MS-SSIM. In \cite{zhang2018learned}, a learned progressive image compression model was proposed using bit-plane decomposition and also bidirectional assembling of gated units. 
Another variable-rate framework was introduced in \cite{akbari2019learned}, which employed GDN-based shortcut connections, stochastic rounding-based scalable quantization, and a variable-rate objective function. The method in \cite{akbari2019learned} outperformed previous learned variable-rate methods.

Most previous works used fixed entropy models shared between the encoder and decoder. In \cite{balle2018variational}, a conditional entropy model based on Gaussian scale mixture (GSM) was proposed where the scale parameters were conditioned on a hyperprior learned using a hyper auto-encoder. The compressed hyperprior was transmitted and added to the bit stream as side information.
This model was extended in \cite{minnen2018joint,lee2018context} where a Gaussian mixture model (GMM) with both mean and scale parameters conditioned on the hyperprior was utilized. In these methods, the hyperpriors were combined with autoregressive priors generated using context models, which outperformed BPG in terms of both PSNR and MS-SSIM. The coding efficiency in \cite{lee2018context} was further improved in \cite{lee2019hybrid} by 
a joint optimization of image
compression and quality enhancement
networks. Another context-adaptive approach was introduced by \cite{zhou2019multi} in which multi-scale masked convolutional networks were utilized for their autoregressive model combined with hyperpriors.

The state of the art in learned image compression has been achieved by context-adaptive entropy methods in which hyperprior and autoregressive models are combined \cite{minnen2018joint}. These approaches are jointly optimized to effectively capture the spatial dependencies and probabilistic structures of the latent representations, which lead to a compression model with superior rate-distortion (R-D) performance. However, similar to natural images,  
the latents are usually represented by
feature maps of the same spatial resolution, which has some spatial redundancies. For example, some of these maps have more low frequency components, and therefore do not need the same resolution as others with more high frequency components. This suggests that better R-D performance can be achieved by having different spatial redundancies in different feature maps.

In  this  paper, a learned multi-frequency (bi-frequency) image compression and entropy model is introduced in which octave convolutions \cite{chen2019drop} are utilized to factorize the latent representations into high frequency (HF) and low frequency (LF) components. The LF information is then represented by a lower spatial resolution, which reduces the corresponding spatial redundancy and improves the compression performance, similar to wavelet transforms \cite{antonini1992image}. In addition, due to the effective communication between HF and LF components in octave convolutions, the reconstruction performance is also improved. In the original octave convolution \cite{chen2019drop}, fixed interpolation methods are used for down- and up-sampling operations, which do not retain the spatial information. So, it can negatively affect the image compression performance. In order to preserve the spatial structure of the latents in our image coding framework, we develop novel generalized octave convolution and octave transposed-convolution architectures with internal activation layers. 

Experiment results show that the proposed scheme outperforms all existing learning-based methods and standard codecs in terms of both PSNR and MS-SSIM on the Kodak dataset.
The framework proposed in this work bridges the wavelet transform and deep learning. Therefore, many techniques in the wavelet transform can be used in the proposed framework to further improve its performance. This will have profound impact on future image coding research. Besides, we show that the proposed generalized octave convolution and transposed-convolution architectures can improve the performance and the computational complexity in other auto-encoder-based computer vision tasks such as semantic segmentation and image denoising.

The paper is organized as follows. In Section \ref{sec:Vanilla vs. Octave Convolution}, vanilla and octave convolutions are briefly described and compared. The proposed generalized octave convolution and transposed-convolution with internal activation layers are formulated and discussed in Section \ref{sec:Proposed Generalized octave Convolution}. The architecture of the proposed multi-frequency image compression framework as well as the multi-frequency entropy model are then introduced in Section \ref{sec:Multi-Frequency Entropy Model}. In Section \ref{sec:experimental}, the experimental results along with the ablation study will then be discussed, and compared with the state-of-the-art in learning-based image compression methods. Following that, the proposed multi-frequency semantic segmentation and image denoising frameworks are respectively studied in Sections \ref{ssec:Multi-Frequency Semantic Segmentation} and \ref{ssec:Multi-Frequency Image Denoising}. The paper conclusion is finally given in Section \ref{concolusion}.

\section{Vanilla vs. Octave Convolution}
\label{sec:Vanilla vs. Octave Convolution}

Let $X,Y \in \mathbb{R}^{h\times w\times c}$ be the input and output feature vectors with $c$ number of channels of size $h\times w$, each feature map in the vanilla convolution is obtained as follows:
\begin{equation}
    Y_{(p,q)}=\sum_{i,j \in \mathcal{N}_k}{\Phi_{(i+\frac{k-1}{2},j+\frac{k-1}{2})}}^{T} X_{(p+i,q+j)},
\end{equation}
where $\Phi \in \mathbb{R}^{k\times k\times c}$ is a $k\times k$ convolution kernel, $(p,q)$ is the location coordinate, and $\mathcal{N}_k$ is a local neighborhood.

In the vanilla convolution, all input and output feature maps are of the same spatial resolution.
If the feature maps in each layer of existing deep learning schemes have the same resolution, there will be some unnecessary redundancies, which will hurt the performance in applications such as compression. For example, some feature maps capture more LF components, and therefore do not need the same resolution as others with more HF components. This suggests that better R-D performance can be achieved by using different spatial redundancies in different feature maps, similar to the approach in wavelet transform \cite{christopoulos2000jpeg2000}.

\begin{figure}[htb!]
 \centering
  \centerline{
  \includegraphics[width=.9\linewidth]{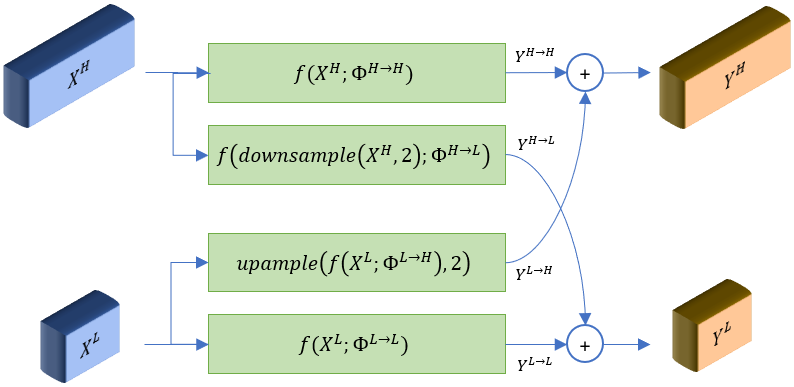}}
\caption{The architecture of the original octave convolution. $X^H$ and $X^L$: input HF and LF feature maps; $f$: regular vanilla convolution; $downsample$: fixed down-sampling operation (e.g., maxpooling); $upsample$: fixed up-sampling operation (e.g., bilinear); $Y^{H\rightarrow H}$ and $Y^{L\rightarrow L}$: intra-frequency updates; $Y^{H\rightarrow L}$ and $Y^{L\rightarrow H}$: inter-frequency communications; $\Phi^{H\rightarrow H}$ and $\Phi^{L\rightarrow L}$: intra-frequency convolution kernels; $\Phi^{H\rightarrow L}$ and $\Phi^{L\rightarrow H}$: inter-frequency convolution kernels; $Y^H$ and $Y^L$: output HF and LF feature maps.}
\label{fig:octconv}
\end{figure}

\begin{figure*}[ht]
\begin{subfigure}{.479\textwidth}
 \centering
  \centerline{
  \includegraphics[width=.9\linewidth]{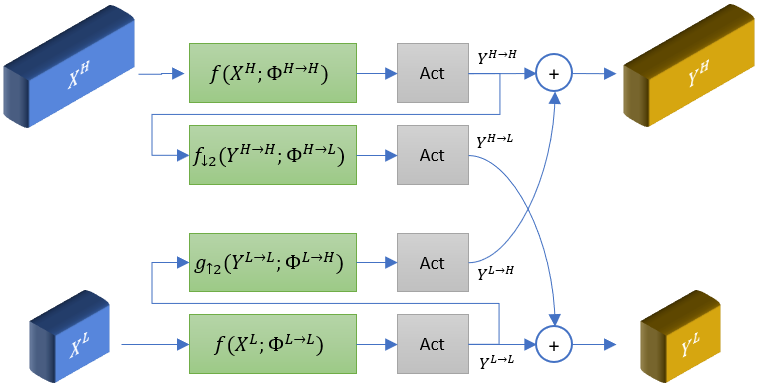}}
  \caption{Architecture of the proposed generalized octave convolution (GoConv). $X^H$ and $X^L$: input HF and LF feature maps; $Y^{H\rightarrow H}$ and $Y^{L\rightarrow L}$: intra-frequency updates; $Y^{H\rightarrow L}$ and $Y^{L\rightarrow H}$: inter-frequency communications; $\Phi^{H\rightarrow H}$ and $\Phi^{L\rightarrow L}$: intra-frequency convolution kernels; $\Phi^{H\rightarrow L}$ and $\Phi^{L\rightarrow H}$: inter-frequency convolution kernels; $Y^H$ and $Y^L$: output HF and LF feature maps.}
\label{fig:goconv}  
\end{subfigure}
\hfill
\begin{subfigure}{.48\textwidth}
 \centering
  \centerline{
  \includegraphics[width=.9\linewidth]{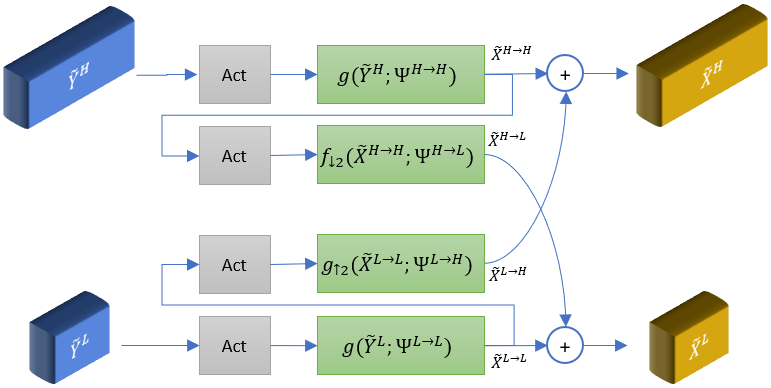}}
  \caption{Architecture of the proposed generalized transposed-convolution (GoTConv). $\Tilde{Y}^H$ and $\Tilde{Y}^L$: input HF and LF feature maps; $\Tilde{X}^{H\rightarrow H}$ and $\Tilde{X}^{L\rightarrow L}$: intra-frequency updates; $\Tilde{X}^{H\rightarrow L}$ and $\Tilde{X}^{L\rightarrow H}$: inter-frequency communications; $\Psi^{H\rightarrow H}$ and $\Psi^{L\rightarrow L}$: intra-frequency transposed-convolution kernels; $\Psi^{H\rightarrow L}$ and $\Psi^{L\rightarrow H}$: inter-frequency convolution kernels; $\Tilde{X}^H$ and $\Tilde{X}^L$: output HF and LF feature maps.}
\label{fig:gotconv}
\end{subfigure}
\caption{Architecture of the proposed generalized octave convolution (GoConv) shown in the left figure, and transposed-convolution (GoTConv) shown in the right figure. \textit{Act}: the activation layer; $f$: regular vanilla convolution; $g$: regular transposed-convolution; $f_{\downarrow2}$: regular convolution with stride 2; $g_{\uparrow2}$: regular transposed-convolution with stride 2.}
\label{fig:conv}
\end{figure*}

To address this problem, in the recently developed octave convolution \cite{chen2019drop}, the feature maps are factorized into HF and LF components with different resolutions, where each component is processed with different convolutions. 
As a result, the resolution of LF feature maps can be spatially reduced, which saves both memory and computation.

The architecture of the original octave convolution is illustrated in Figure \ref{fig:octconv}. The factorization of input vector $X$ in octave convolutions is denoted by $X = \{X^H, X^L\}$, where $X^H \in \mathbb{R}^{h\times w \times(1-\alpha)c}$ and $X^L \in \mathbb{R}^{\frac{h}{2} \times \frac{w}{2} \times \alpha c}$ are respectively the HF and LF maps. The ratio of channels allocated to the LF feature representations (i.e., at half of spatial resolution) is defined by $\alpha \in [0,1]$. 
The factorized output vector is denoted by $Y = \{Y^H, Y^L\}$, where $Y^H \in \mathbb{R}^{h'\times w' \times(1-\alpha)c'}$ and $Y^L \in \mathbb{R}^{\frac{h'}{2} \times \frac{w'}{2} \times \alpha c'}$ are the output HF and LF maps. The outputs are given by:
\begin{equation}
\begin{split}
Y^H = Y^{H\rightarrow H} + Y^{L\rightarrow H},\\
Y^L = Y^{L\rightarrow L} + Y^{H\rightarrow L},\\
\end{split}
\end{equation}
where $Y^{H\rightarrow H}$ and $Y^{L\rightarrow L}$ are intra-frequency update and $Y^{H\rightarrow L}$ and $Y^{L\rightarrow H}$ denote inter-frequency communication. intra-frequency component is used to update the information within each part, while inter-frequency communication further enables information exchange between the two parts. Similar to filter bank theory \cite{vaidyanathan2006multirate}, the octave convolution allows information exchange between the HF and LF feature maps.

The octave convolution kernel is given by $\Phi = [\Phi^H, \Phi^L]$ with which the inputs $X^H$ and $X^L$ are respectively convolved. $\Phi^H$ and $\Phi^L$ are further divided into intra- and inter-frequency components as follows: $\Phi^H = [\Phi^{H\rightarrow H}, \Phi^{L\rightarrow H}]$ and $\Phi^L = [\Phi^{L\rightarrow L}, \Phi^{H\rightarrow L}]$. 

For the intra-frequency update, the regular vanilla convolution is used. However, up- and down-sampling interpolations are applied to compute the inter-frequency communication formulated as:
\begin{equation}
\begin{split}
       Y^{H} &= f(X^H;\Phi^{H\rightarrow H})+upsample\left(f(X^L;\Phi^{L\rightarrow H}),2\right), \\
       Y^{L} &= f(X^L;\Phi^{L\rightarrow L})+f\left(downsample(X^H,2);\Phi^{H\rightarrow L}\right),    
\end{split}
\end{equation}
where $f$ denotes a vanilla convolution with parameters $\Phi$.

As reported in \cite{chen2019drop}, due to the effective inter-frequency communications, the octave convolution can have better performance in classification and recognition performance compared to the vanilla convolution. 
Since the octave convolution allows different resolutions in HF and LF feature maps, it is very suitable for image compression. This motivates us to apply it to learned image compression. However, some modifications are necessary in order to get a good performance.

\section{Generalized Octave Convolution}
\label{sec:Proposed Generalized octave Convolution}

In the original octave convolution, the average pooling and nearest interpolation are respectively employed for down- and up-sampling operations in inter-frequency communication \cite{chen2019drop}. Such conventional interpolations 
do not preserve spatial information and structure of the input feature map.
In addition, in convolutional auto-encoders where sub-sampling needs to be reversed at the decoder side, fixed operations such as pooling result in a poor performance \cite{springenberg2014striving}.

In this work, we propose a novel generalized octave convolution (GoConv) in which strided convolutions are used to sub-sample the feature vectors and compute the inter-frequency communication in a more effective way. Fixed sub-sampling operations such as pooling are designed to forget about spatial structure, for example, in object recognition where we only care about the presence or absence of the object, not its position. However, if the spatial information is important, strided convolution can be a useful alternative. With learned filters, strided convolutions can learn to handle discontinuities from striding and preserve more spatial properties required in down-sampling operation \cite{springenberg2014striving}. Moreover, since it can learn how to summarize, better generalization with respect to the input is achieved. As a result, better performance with less spatial information loss can be achieved, especially in auto-encoders where it is easier to reverse strided convolutions. Moreover, as in ResNet, applying strided convolution (i.e., convolution and down-sampling at the same time) reduces the computational cost compared to a convolution followed by a fixed down-sampling operation (e.g., average pooling).

\begin{figure*}
 \centering
  \centerline{\includegraphics[width=\textwidth]{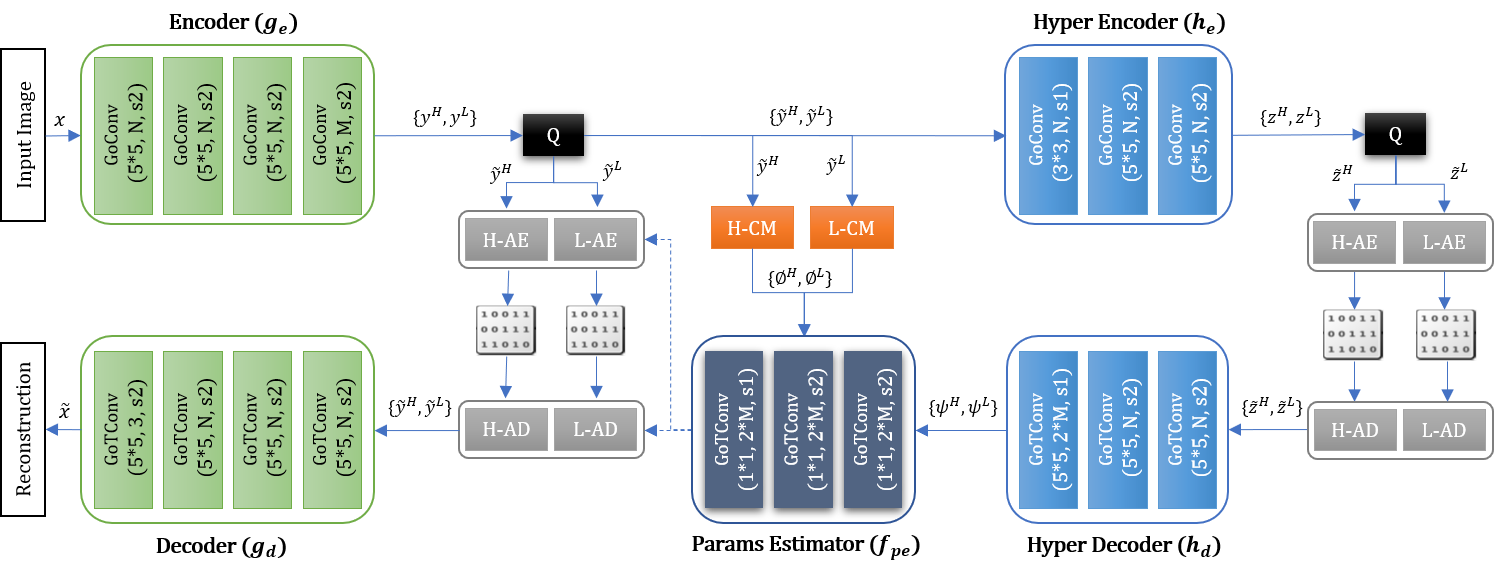}}
\caption{Overall framework of the proposed image compression model. \textbf{H-AE} and \textbf{H-AD}: arithmetic encoder and decoder for HF latents. \textbf{L-AE} and \textbf{L-AD}: arithmetic encoder and decoder for LF latents. \textbf{H-CM} and \textbf{L-CM}: the HF and LF context models each composed of one 5*5 masked convolution layer with 2*M filters and stride of 1. \textbf{Q}: represents the additive uniform noise for training, or uniform quantizer for the test.}
\label{fig:framework}
\end{figure*} 

The architecture of the proposed GoConv is shown in Figure \ref{fig:goconv}. Compared to the original octave convolution (Figure \ref{fig:octconv}), we apply another important modification regarding the inputs to the inter-frequency convolution operations. As summarized in Section \ref{sec:Vanilla vs. Octave Convolution}, in order to calculate the inter-frequency communication outputs (denoted by $Y^{H\rightarrow L}$ and $Y^{L\rightarrow H}$), the input HF and LF vectors (denoted by $X^H$ and $X^L$) are considered as the inputs to the HF-to-LF and LF-to-HF convolutions, respectively ($f_{\downarrow2}$ and $g_{\uparrow2}$ in Figure \ref{fig:goconv}). This strategy is only efficient for the stride of 1 (i.e., the size of input and output HF and LF vectors is the same). However, in GoConv, this procedure can result in significant information loss for larger strides. As an example, consider using stride 2, which results in down-sampled output HF and LF feature maps (half resolution of input HF and LF maps). To achieve this, stride 2 is required for the intra-frequency convolution ($f$ in Figure \ref{fig:goconv}). However, for the inter-frequency convolution $f_{\downarrow2}$, a harsh stride of 4 should be used, which results in significant spatial information loss. 

In order to deal with this problem, we instead use two consecutive convolutions with stride 2 where the first convolution is indeed the intra-frequency operation $f$. In other words, to compute $Y^{H\rightarrow L}$, we exploited the filters learned by $f$ to have less information loss. Thus, instead of $X^H$ and $X^L$, we set $Y^{H\rightarrow H}$ and $Y^{L\rightarrow L}$ as inputs to $f_{\downarrow2}$ and $g_{\uparrow2}$. 

The output HF and LF feature maps in GoConv are formulated as follows:
\begin{equation}
\begin{split}
       Y^{H} = Y^{H\rightarrow H}+g_{\uparrow2}(Y^{L\rightarrow L};\Phi^{L\rightarrow H}), \\
       Y^{L} = Y^{L\rightarrow L}+f_{\downarrow2}(Y^{H\rightarrow H};\Phi^{H\rightarrow L}), \\
      \text{ with }  
      Y^{H\rightarrow H} =f(X^H;\Phi^{H\rightarrow H}),\\
      Y^{L\rightarrow L}=f(X^L;\Phi^{L\rightarrow L}),
\end{split}
\label{equation:GoConv}
\end{equation}
where $f_{\downarrow2}$ and $g_{\uparrow2}$ are respectively Vanilla convolution and transposed-convolution operations with stride of 2.

In original octave convolutions, activation layers (e.g., ReLU) are applied to the output HF and LF maps. However, as shown in Figure \ref{fig:conv}, we utilize activations for each internal convolution performed in our proposed GoConv. In this case, we assure activation functions are properly applied to each feature map computed by convolution operations. Each of the inter- and intra-frequency components is then followed by an activation layer in GoConv. 

We also propose a generalized octave transposed-convolution denoted by GoTConv (Figure \ref{fig:conv}), which can replace the conventional transposed-convolution commonly employed in deep auto-encoder (encoder-decoder) architectures. Let $\Tilde{Y} = \{\Tilde{Y}^{H}, \Tilde{Y}^{L}\}$ and $\Tilde{X} = \{\Tilde{X}^{H}, \Tilde{X}^{L}\}$ respectively be the factorized input and output feature vectors, the output HF and LF maps $\Tilde{X}^{H}$ and $\Tilde{X}^{L}$) in GoTConv are obtained as follows:
\begin{equation}
\begin{split}
       \Tilde{X}^{H} = \Tilde{X}^{H\rightarrow H}+g_{\uparrow2}(\Tilde{X}^{L\rightarrow L};\Psi^{L\rightarrow H}), \\
       \Tilde{X}^{L} = \Tilde{X}^{L\rightarrow L}+f_{\downarrow2}(\Tilde{X}^{H\rightarrow H};\Psi^{H\rightarrow L}) \\
      \text{ with }      \Tilde{X}^{H\rightarrow H} = g(\Tilde{Y}^H;\Psi^{H\rightarrow H}),\\
      \Tilde{X}^{L\rightarrow L} = g(\Tilde{Y}^L;\Psi^{L\rightarrow L}),
\end{split}
\label{equation:GoTConv}
\end{equation}
where $\Tilde{Y}^H, \Tilde{X}^H \in \mathbb{R}^{h\times w \times(1-\alpha)c}$ and $\Tilde{Y}^L, \Tilde{X}^L \in \mathbb{R}^{\frac{h}{2} \times \frac{w}{2} \times \alpha c}$. Unlike GoConv in which regular convolution operation is used, transposed-convolution denoted by $g$ is applied for intra-frequency update in GoTConv. For up- and down-sampling operations in inter-frequency communication, the same strided convolutions $g_{\uparrow2}$ and $f_{\downarrow2}$ as in GoConv are respectively utilized.

Similar to the original octave convolution, the proposed GoConv and GoTConv are designed and formulated as generic, plug-and-play units. As a result, they can respectively replace vanilla convolution and transposed-convolution units in any convolutional neural network (CNN) architecture, especially auto-encoder-based frameworks such as image compression, image denoising, and semantic segmentation. When used in an auto-encoder, the input image to the encoder is not represented as a multi-frequency tensor. In this case, to compute the output of the first GoConv layer in the encoder, Equation \ref{equation:GoConv} is modified as follows:
\begin{equation}
       Y^{H} = f(X;\Phi^{H\rightarrow H}),~~~Y^{L} = f_{\downarrow2}(Y^{H};\Phi^{H\rightarrow L}),
\label{equation:GoConv2}
\end{equation}

Similarly, at the decoder side, the output of the last GoTConv is a single tensor representation, which can be formulated by modifying Equation \ref{equation:GoTConv} as:
\begin{equation}
\begin{split}
       \Tilde{X} = \Tilde{X}^{H\rightarrow H}+g_{\uparrow2}(\Tilde{X}^{L\rightarrow L};\Psi^{L\rightarrow H}), \\
      \text{ with }      \Tilde{X}^{H\rightarrow H} = g(\Tilde{Y}^H;\Psi^{H\rightarrow H}),\\
      \Tilde{X}^{L\rightarrow L} = g(\Tilde{Y}^L;\Psi^{L\rightarrow L}).
\end{split}
\label{equation:GoTConv2}
\end{equation}

Compared to GoConv, the process of using activations for each internal transposed-convolution in GoTConv is inverted, where the activation layer is followed by inter- and intra-frequency communications as shown in Figure \ref{fig:conv}.

\section{Multi-Frequency Image Coding and Entropy Model}
\label{sec:Multi-Frequency Entropy Model}

Octave convolution is similar to the wavelet transform \cite{antonini1992image}, since it has lower resolution in LF than in HF. Therefore, it can be used to improve the R-D performance in learning-based image compression frameworks. 
Moreover, due to the effective inter-frequency communication as well as the receptive field enlargement in octave convolutions, they also improve the performance of the analysis (encoding) and synthesis (decoding) transforms in a compression framework.

The overall architecture of the proposed multi-frequency image compression framework is shown in Figure \ref{fig:framework}. Similar to \cite{minnen2018joint}, our architecture is composed of two sub-networks: the core auto-encoder and the entropy sub-network. The core auto-encoder is used to learn a quantized latent vector of the input image, while the entropy sub-network is responsible for learning a probabilistic model over the quantized latent representations, which is utilized for entropy coding.

In order to handle multi-frequency entropy coding, we have made several improvements to the scheme in \cite{minnen2018joint}. First, all vanilla convolutions in the core encoder, and hyper encoder are replaced by the proposed GoConv, and all vanilla transposed-convolutions in the core and hyper decoders are replaced by GoTConv. In \cite{minnen2018joint}, each convolution/transposed-convolution is accompanied by an activation layer (e.g., GDN/IGDN or Leaky ReLU). In our architecture, we move these layers into the GoConv and GoTConv architectures and directly apply them to the inter- and intra-frequency components as described in Section \ref{sec:Proposed Generalized octave Convolution}. GDN/IGDN transforms are respectively used for the GoConv and GoTConv employed in the proposed deep encoder and decoder, while Leaky ReLU is utilized for the hyper auto-encoder and the parameters estimator. The convolution properties (i.e., size and number of filters and strides) of all networks including the core and hyper auto-encoders, context models, and parameter estimator are the same as in \cite{minnen2018joint}.

Let $x \in \mathbb{R}^{h\times w\times 3}$ be the input image, the multi-frequency latent representations are denoted by $\{y^{H},y^{L}\}$ where $y^{H} \in \mathbb{R}^{\frac{h}{16} \times \frac{w}{16} \times (1-\alpha)M}$ and $y^{L} \in \mathbb{R}^{\frac{h}{32} \times \frac{w}{32} \times \alpha M}$ are generated using the parametric deep encoder (i.e., analysis transform) $g_e$ represented as:
\begin{equation}
\{y^{H},y^{L}\}=g_e(x;\theta_{ge}),
\end{equation}
where $\theta_{ge}$ is the parameter vector to be optimized. $M$ denotes the total number of output channels in $g_e$, which is divided into $(1-\alpha)M$ channels for HF and $\alpha M$ channels for LF (i.e., at half spatial resolution of the HF part). The calculation in Equation (\ref{equation:GoConv2}) is used for the first GoConv layer, while the other encoder layers are formulated using Equation (\ref{equation:GoConv}).

At the decoder side, 
the parametric decoder (i.e., synthesis transform) $g_d$ with the parameter vector $\theta_{gd}$ reconstructs the image $\Tilde{x} \in \mathbb{R}^{h\times w\times 3}$ as follows: 
\begin{equation}
\begin{split}
    \Tilde{x}=g_d\left(\{\Tilde{y}^{H},\Tilde{y}^{L}\};\theta_{gd}\right)
    \text{  with  } 
    \{\Tilde{y}^{H},\Tilde{y}^{L}\} &= Q\left(\{y^H,y^L\}\right),
\end{split}
\end{equation}
where $Q$ represents the addition of uniform noise to the latent representations during training, or uniform quantization (i.e., round function in this work) and arithmetic coding/decoding of the latents during the test. As illustrated in Figure \ref{fig:framework}, the quantized HF and LF latents $\Tilde{y}^{H}$ and $\Tilde{y}^{L}$ are entropy-coded using two separate arithmetic encoder and decoder.

The entropy sub-network in our architecture contains two models: a context model and a hyper auto-encoder \cite{minnen2018joint}. The context model is an autoregressive model over multi-frequency latent representations. Unlike the other networks in our architecture where GoConv are incorporated for their convolutions, we use Vanilla convolutions in the context model to ensure that the causality of the contexts is not spoiled due to the intra-frequency communication in GoConv. The contexts of the HF and LF latents, denoted by $\phi^H_i$ and $\phi^L_i$, are then predicted with two separate models $f^H_{cm}$ and $f^L_{cm}$ defined as follows:
\begin{equation}
\begin{split}
    \phi^H_i=f^H_{cm}(\Tilde{y}^H_{<i};\theta^H_{cm}) \text{ and } \phi^L_i=f^L_{cm}(\Tilde{y}^L_{<i};\theta^L_{cm}),
\end{split}
\end{equation}
where $\theta^H_{cm}$ and $\theta^L_{cm}$ are the parameters to be generalized. Both $f^H_{cm}$ and $f^L_{cm}$ are composed of one 5*5 masked convolution \cite{van2016conditional} with stride of 1.

The hyper auto-encoder learns to represent side information useful for correcting the context-based predictions. The spatial dependencies of $\{\Tilde{y}^{H},\Tilde{y}^{L}\}$ are then captured into the multi-frequency hyper latent representations $\{z^{H},z^{L}\}$ using the parametric hyper encoder $h_e$ (with the parameter vector $\theta_{he}$) defined as:
\begin{equation}
\begin{split}
    \{z^{H},z^{L}\}=h_e\left(\{\Tilde{y}^{H},\Tilde{y}^{L}\};\theta_{he}\right).
\end{split}    
\end{equation}

\begin{figure}
\centering
\begin{subfigure}[b]{.24\textwidth}
 \centering
  \centerline{\includegraphics[width=\textwidth]{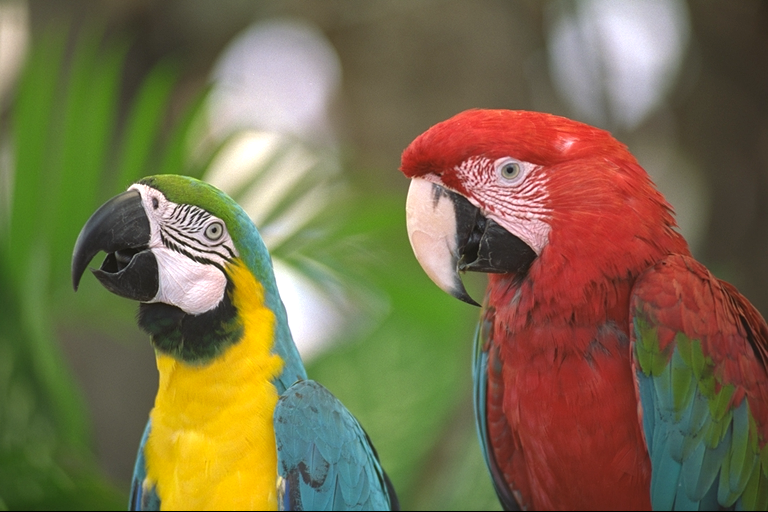}}
\end{subfigure}
\begin{subfigure}[b]{.079\textwidth}
 \centering
  \centerline{\includegraphics[width=\textwidth]{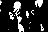}}
  \centerline{\includegraphics[width=\textwidth]{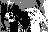}}
  \centerline{\includegraphics[width=\textwidth]{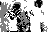}}
\end{subfigure}
\begin{subfigure}[b]{.079\textwidth}
 \centering
    \centerline{\includegraphics[width=\textwidth]{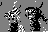}}
  \centerline{\includegraphics[width=\textwidth]{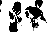}}
  \centerline{\includegraphics[width=\textwidth]{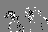}} 
\end{subfigure}
\begin{subfigure}[b]{.070\textwidth}
 \centering
    \centerline{\includegraphics[width=0.85cm]{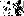}}
    \centerline{\includegraphics[width=0.85cm]{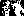}}  
    \centerline{\includegraphics[width=0.85cm]{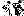}}  
    \centerline{\includegraphics[width=0.85cm]{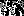}}  
    \centerline{\includegraphics[width=0.85cm]{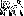}}   
\end{subfigure}
\caption{Sample HF and LF latent representations. Left column: original image; Middle columns: HF; Right column: LF.}
\label{fig:features}
\end{figure}

\begin{figure*}
\begin{minipage}{0.49\linewidth}
 \centering
  \centerline{\includegraphics[width=\textwidth]{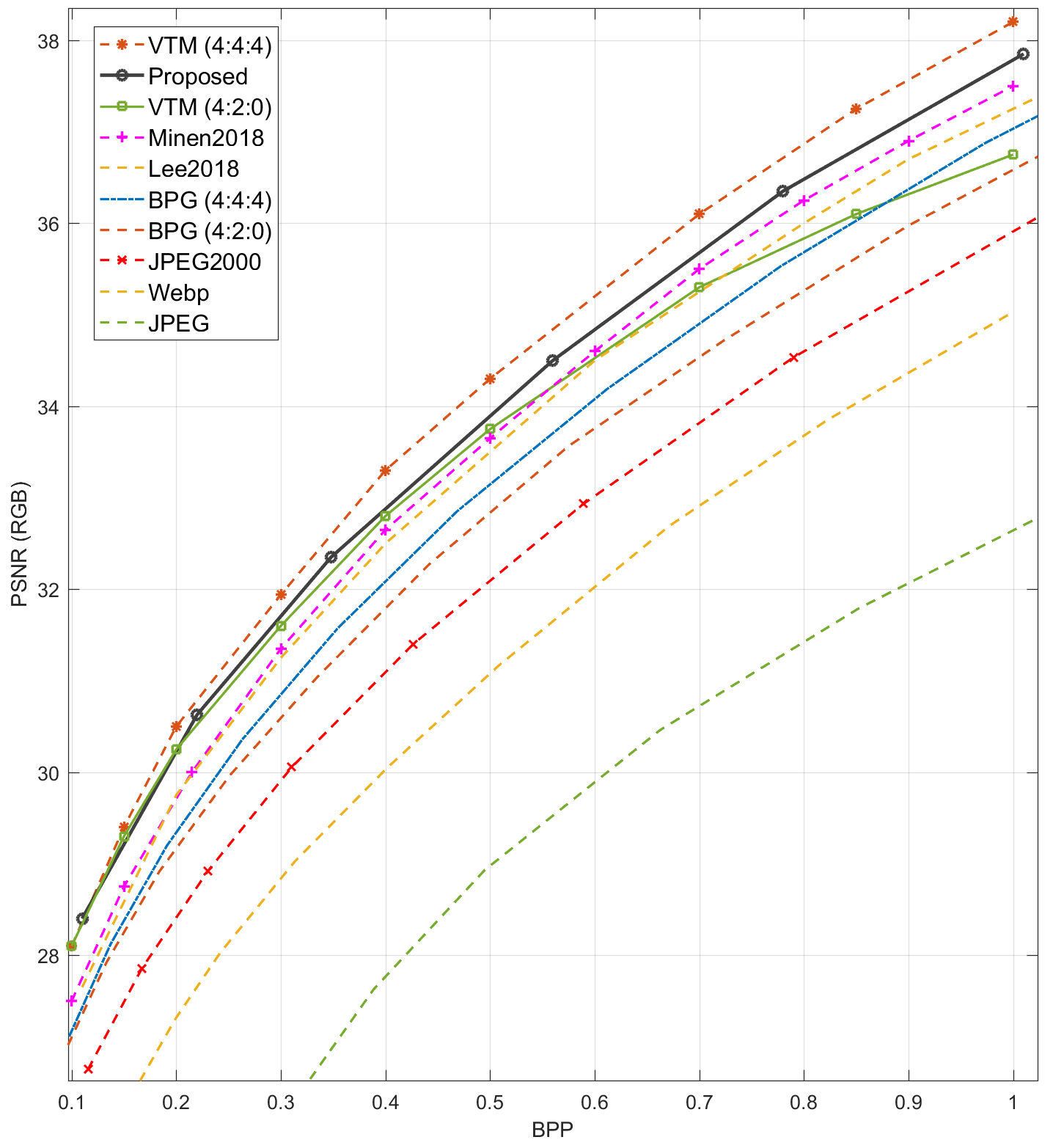}}
\end{minipage}
\begin{minipage}{0.49\linewidth}
 \centering
  \centerline{\includegraphics[width=\textwidth]{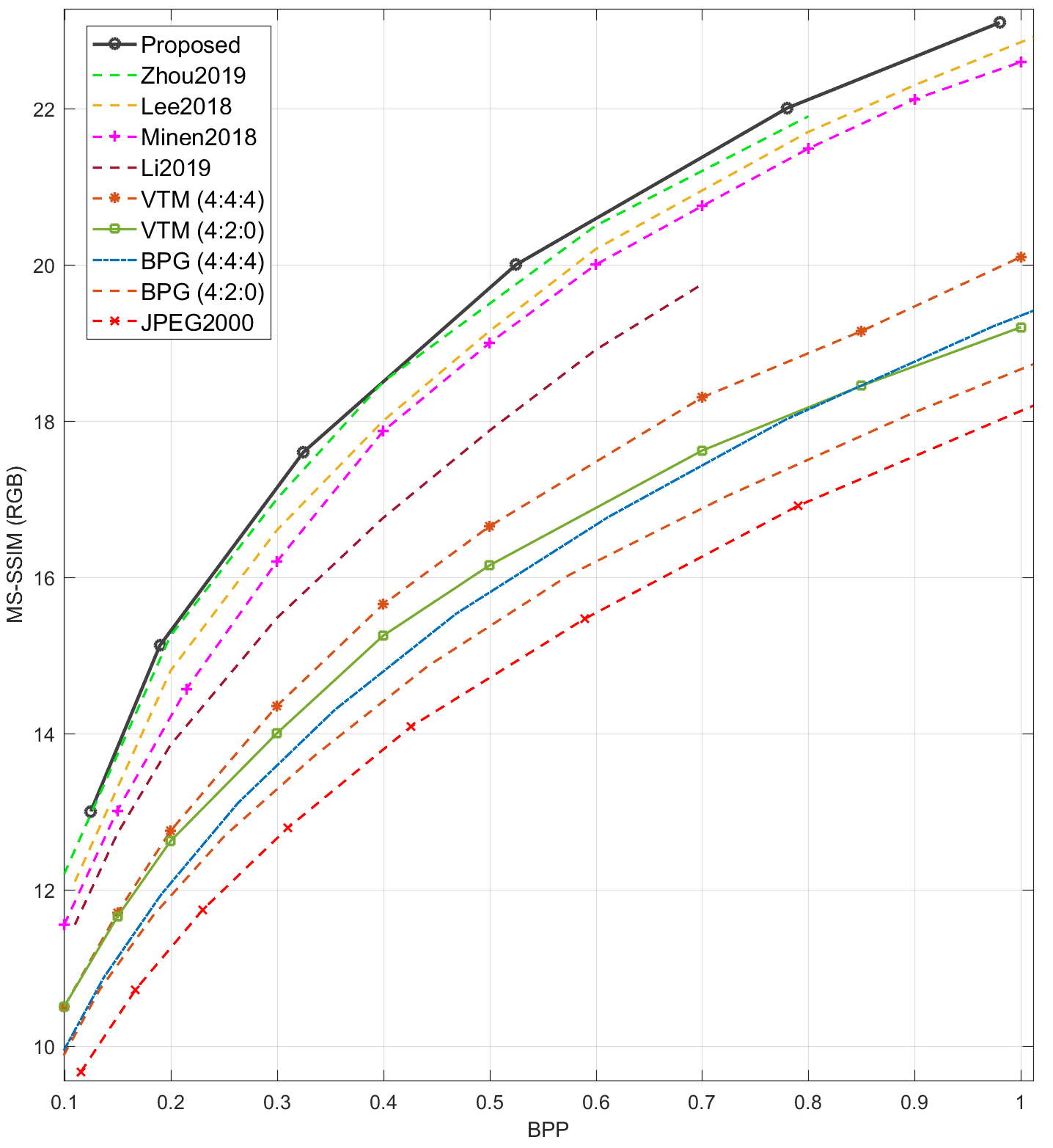}}
\end{minipage}
\caption{Kodak comparison results of our approach with traditional codecs and learning-based image compression methods.}
\label{fig:results_Kodak}
\end{figure*}

The quantized hyper latents are also part of the generated bitstream that is required to be entropy-coded and transmitted. Similar to the core latents, two separate arithmetic coders are used for the quantized HF and LF $\Tilde{z}^H$ and $\Tilde{z}^L$. Given the quantized hyper latents, the side information used for the entropy model estimation is reconstructed using the hyper decoder $h_d$ (with the parameter vector $\theta_{hd}$) formulated as:
\begin{equation}
\begin{split}    
    \{\psi^H,\psi^L\}=h_d\left(\{\Tilde{z}^H,\Tilde{z}^L\};\theta_{hd}\right)\\
    \text{  with  } 
    \{\Tilde{z}^{H},\Tilde{z}^{L}\} = Q\left(\{z^H,z^L\}\right).
\end{split}    
\end{equation}

As shown in Figure \ref{fig:framework}, to estimate the mean and scale parameters required for a conditional Gaussian entropy model, the information from both context model and hyper decoder is combined by another networks, denoted by $f^H_{pe}$ and $f^L_{pe}$ (with the parameter vectors $\theta^H_{ep}$ and $\theta^L_{ep}$), represented as follows: 
\begin{equation}
    \{\mu^H_i,\sigma^H_i\}=
    f^H_{pe}\left(\{\psi^H,\phi^H_i\};\theta^H_{ep}\right),
\label{equation:parameters estimator high}
\end{equation}
\begin{equation}
    \{\mu^L_i,\sigma^L_i\}=
    f^L_{pe}\left(\{\psi^L,\phi^L_i\};\theta^L_{ep}\right),
\label{equation:parameters estimator low}
\end{equation}
where $\mu^H_i$ and $\sigma^H_i$ are the parameters for entropy modelling of the HF information, and  $\mu^L_i$ and $\sigma^L_i$ are for the LF information.

The objective function for training is composed of two terms: rate $R$, which is the expected length of the bitstream, and distortion $D$, which is the expected error between the input and reconstructed images. The R-D balance is determined by a Lagrange multiplier denoted by $\lambda$. The R-D optimization problem is then defined as follows:
\begin{equation}
\begin{split}
    \mathcal{L}=R+\lambda D\\ \text{  with  }    
    R&=R^H+R^L,\\
    D&=\mathbb{E}_{x	\sim p_x}\left[d(x,\hat{x})\right],
\end{split}
\label{equation:R-D}
\end{equation}
where $p_x$ is the unknown distribution of natural images and $D$ can be any distortion metric such as mean squared error (MSE) or MS-SSIM. $R^H$ and $R^L$ are the rates corresponding to the HF and LF information (bitstreams) defined as follows:
\begin{equation}
\begin{split}
    R^H=\mathbb{E}_{x	\sim p_x}\left[-\log_2 p_{\Tilde{y}^H|\Tilde{z}^H}(\Tilde{y}^H|\Tilde{z}^H) \right]\\
    + \mathbb{E}_{x	\sim p_x}\left[-\log_2 p_{\Tilde{z}^H}(\Tilde{z}^H) \right],\\
    R^L=\mathbb{E}_{x	\sim p_x}\left[-\log_2 p_{\Tilde{y}^L|\Tilde{z}^L}(\Tilde{y}^L|\Tilde{z}^L) \right]\\
    + \mathbb{E}_{x	\sim p_x}\left[-\log_2 p_{\Tilde{z}^L}(\Tilde{z}^L) \right],\\
\end{split}    
\label{equation:rate}
\end{equation}
where $p_{\Tilde{y}^H|\Tilde{z}^H}$ and $p_{\Tilde{y}^L|\Tilde{z}^L}$ are respectively the conditional Gaussian entropy models for HF and LF latent representations ($y^H$ and $y^L$) formulated as:
\begin{equation}
\begin{split}
    p_{\Tilde{y}^H|\Tilde{z}^H}(\Tilde{y}^H|\Tilde{z}^H,\theta_{hd},\theta^H_{cm},\theta_{ep})=\\
    \prod_{i}\left(\mathcal{N}(\mu^H_i,{\sigma_i^{\scaleto{2}{3pt}}}^H)*\mathcal{U}(-\tfrac{1}{2},\tfrac{1}{2}) \right)(\Tilde{y}^H_i),\\
    p_{\Tilde{y}^L|\Tilde{z}^L}(\Tilde{y}^L|\Tilde{z}^L,\theta_{hd},\theta^L_{cm},\theta_{ep})=\\
    \prod_{i}\left(\mathcal{N}(\mu^L_i,{\sigma_i^{\scaleto{2}{3pt}}}^L)*\mathcal{U}(-\tfrac{1}{2},\tfrac{1}{2}) \right)(\Tilde{y}^L_i),\\
\end{split}
\end{equation}
where each latent is modelled as a Gaussian convolved with a unit uniform distribution, which ensures a good match between encoder and decoder distributions of both quantized and continuous-values latents. The mean and scale parameters $\mu^H_i$, $\sigma^H_i$, $\mu^L_i$, and $\sigma^L_i$ are generated via the networks $f^H_{pe}$ and $f^L_{pe}$ defined in Equations \ref{equation:parameters estimator high} and \ref{equation:parameters estimator low}.

Since the compressed hyper latents $\Tilde{z}^H$ and $\Tilde{z}^L$ are also part of the generated bitstream, their transmission costs are also considered in the rate term formulated in Equation \ref{equation:rate}. As in \cite{balle2018variational,minnen2018joint}, to model HF and LF hyper-priors, we assume the entries to be independent and identically distributed (i.i.d.) and fit a univariate piecewise linear density model to represent each channel $j$. The non-parametric, fully-factorized density models for the HF and LF hyper latents are then formulated as follows:
\begin{equation}
\begin{split}
    p_{\Tilde{z}^H|\Theta^H}(\Tilde{z}^H|\Theta^H)=\prod_{j}\left(p_{{z^H_i}|\Theta^H_j}(\Theta^H_j)*\mathcal{U}(-\tfrac{1}{2},\tfrac{1}{2}) \right)(\Tilde{z}^H_j),\\
    p_{\Tilde{z}^L|\Theta^L}(\Tilde{z}^L|\Theta^L)=\prod_{j}\left(p_{{z^L_i}|\Theta^L_j}(\Theta^L_j)*\mathcal{U}(-\tfrac{1}{2},\tfrac{1}{2}) \right)(\Tilde{z}^H_j),
\end{split}    
\end{equation}
where $\Theta^H$ and $\Theta^L$ denote the parameter vectors for the univariate distributions $p_{\Tilde{z}^H|\Theta^H}$ and $p_{\Tilde{z}^L|\Theta^L}$.

\begin{figure*}
\centering
\begin{subfigure}[b]{.24\textwidth}
 \centering
  \centerline{\includegraphics[width=\textwidth]{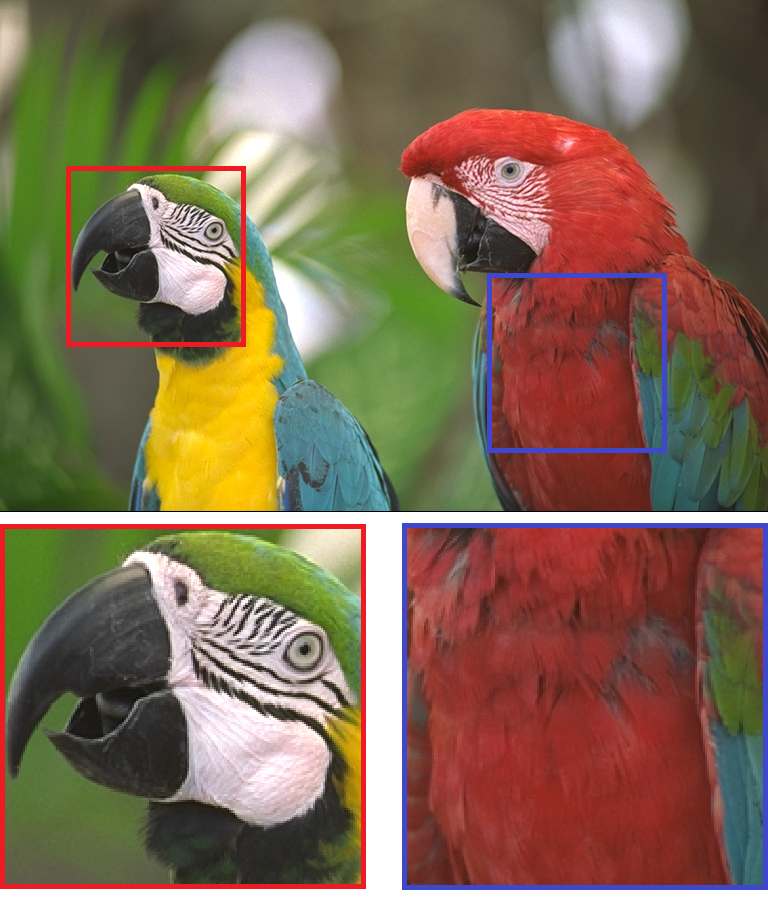}}
  \subcaption{\scriptsize Original}
\end{subfigure}
\begin{subfigure}[b]{.24\textwidth}
 \centering
  \centerline{\includegraphics[width=\textwidth]{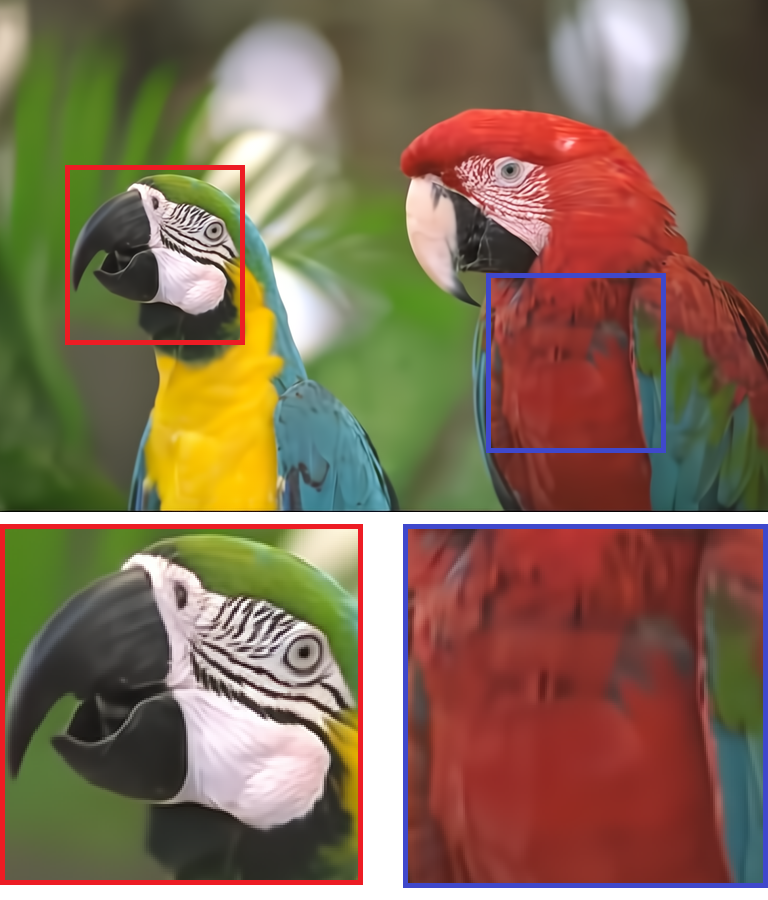}}
  \subcaption{\scriptsize Ours (0.149bpp, 35.30dB, 15.25dB)}
\end{subfigure}
\begin{subfigure}[b]{.24\textwidth}
 \centering
  \centerline{\includegraphics[width=\textwidth]{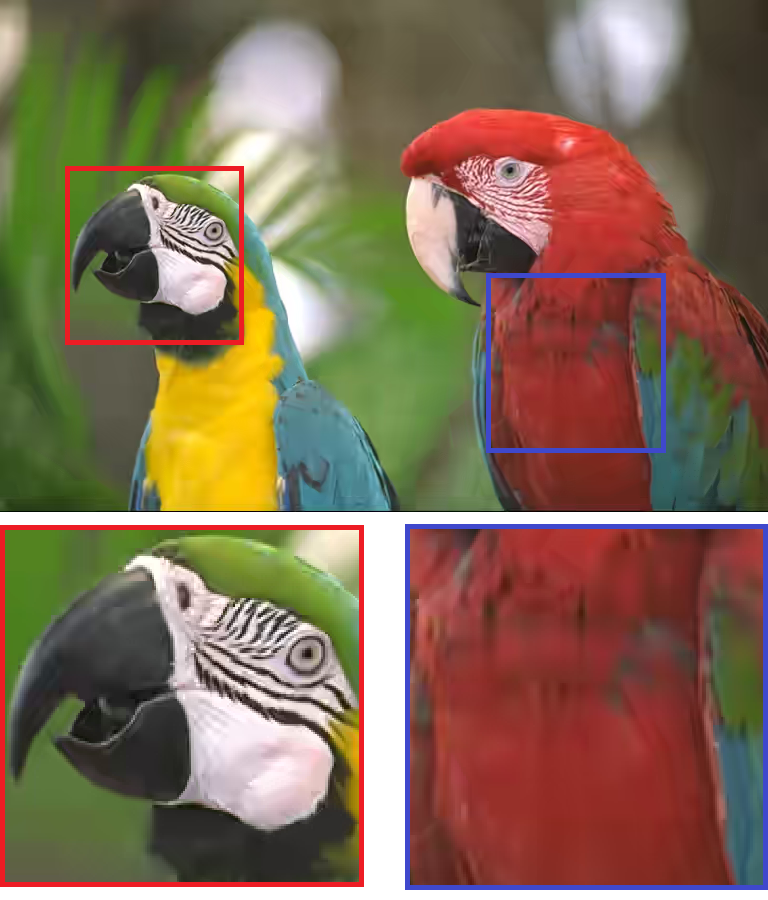}}
  \subcaption{\scriptsize BPG (0.151bpp, 33.96dB, 14.44dB)}
\end{subfigure}
\begin{subfigure}[b]{.24\textwidth}
 \centering
  \centerline{\includegraphics[width=\textwidth]{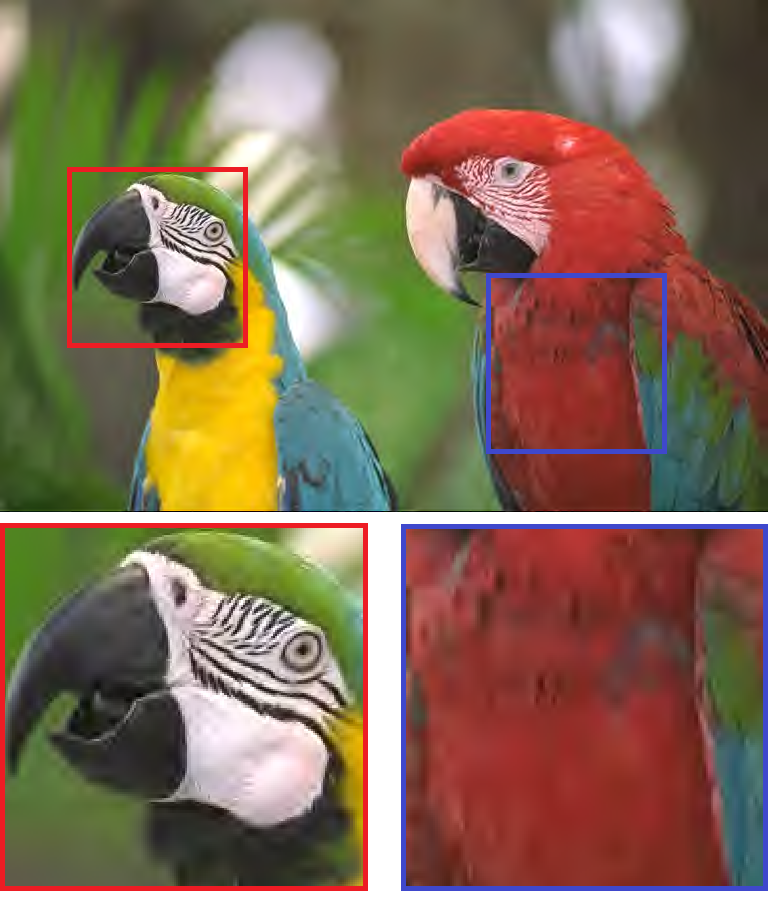}}
  \subcaption{\scriptsize J2K (0.151bpp, 32.59dB, 13.43dB)}
\end{subfigure}
\caption{Kodak visual example (bits-per-pixel, PSNR, $\text{MS-SSIM}_\text{dB}$). 
}
\label{fig:kodak_quan1}
\end{figure*}

\section{Experimental Results: Image Compression}
\label{sec:experimental}

The CLIC training set with images of at least 256 pixels in height or width (1732 images in total) were used for training the proposed model. Random crops of size $h=w=256$ were extracted from the images for training. We set $\alpha=0.5$ so that 50\% of the latent representations is assigned to the LF part with half spatial resolution. Sample HF and LF latent representations are shown in Figure \ref{fig:features}.

Considering four layers of strided convolutions (with stride of 2) and the output channel size $M=192$ in the core encoder (Figure \ref{fig:framework}), the HF and LF latents $y^H$ and $y^L$ will respectively be of size 16$\times$16$\times$96 and 8$\times$8$\times$96 for training. As discussed in \cite{balle2018variational}, 
the optimal number of filters (i.e., $N$) increases with the R-D balance factor $\lambda$, which indicates that higher network capacity is required for models with higher bit rates. As a result, in order to avoid $\lambda$-dependent performance saturation and to boost the network capacity, we set $M=N=256$ for higher bit rates (BPPs > 0.5). All models in our framework were jointly trained for 200 epochs with mini-batch stochastic gradient descent and a batch size of 8. The Adam solver with learning rate of 0.00005 was fixed for the first 100 epochs, and was gradually decreased to zero for the next 100 epochs. 

We compare the performance of the proposed scheme with standard codecs including JPEG, JPEG2000 \cite{christopoulos2000jpeg2000}, WebP \cite{webp2018}, BPG (both YUV4:2:0 and YUV4:4:4 formats) \cite{bellard2017bpg}, the next-generation video coding standard VVC Test Model or VTM 5.2 (both YUV4:2:0 and YUV4:4:4 formats) \cite{vvc-vtm}, and also state-of-the-art learned image compression methods in \cite{minnen2018joint, li2019learning, lee2018context, lee2019hybrid, zhou2019multi}. We use both PSNR and $\text{MS-SSIM}_\text{dB}$ as the evaluation metrics, where $\text{MS-SSIM}_\text{dB}$ represents MS-SSIM scores in dB defined as: $\text{MS-SSIM}_\text{dB}=-10 log_{10}(1-\text{MS-SSIM})$. 

The comparison results on the popular Kodak image set (averaged over 24 test images) are shown in Figure \ref{fig:results_Kodak}. For the PSNR results, we optimized the model for the MSE loss as the distortion metric $d$ in Equation \ref{equation:R-D}, while the perceptual MS-SSIM metric was used for the MS-SSIM results reported in Figure \ref{fig:results_Kodak}. In order to obtain the seven different bit rates on the R-D curve illustrated in Figure \ref{fig:results_Kodak}, seven models with seven different values for $\lambda$ were trained.

As shown in Figure \ref{fig:results_Kodak}, our method outperforms the standard codecs such as BPG and VTM (4:2:0) as well as the state-of-the-art learning-based image compression methods in terms of both PSNR and MS-SSIM. Our method achieves $\approx$0.25dB lower PSNR than VTM (4:4:4). However, compared to VTM (4:2:0), the proposed approach provides $\approx$0.12dB better PSNR at lower bit rates (bpp < 0.5) and $\approx$0.5-1dB better PSNR at higher rates.

\begin{table*}
\caption{Ablation study of different components in the proposed framework. \textbf{BPP}: bits-per-pixel (HF/LF: BPPs for HF and LF latents). \textbf{ActOut}: activation layers moved out of GoConv/GoTConv; \textbf{CoreOct}: proposed GoConv/GoTConv only used for the core auto-encoder; \textbf{OrgOct}: GoConv/GoTConv replaced by original octave convolutions.}
\centering
\begin{tabular}{|c|c|c|c|c|c|c|} 
\cline{2-7}
\multicolumn{1}{l|}{}     &
\textbf{$\boldsymbol{\alpha}$ = 0.25}  & \textbf{$\boldsymbol{\alpha}$ = 0.5}  & \textbf{$\boldsymbol{\alpha}$ = 0.75}  & \textbf{ActOut}  & \textbf{CoreOct}  & \textbf{OrgOct}   \\ 
\hhline{|-|=|=|=|=|=|=|}
 \begin{tabular}[c]{@{}c@{}} \textbf{BPP}\\\textbf{(HF~~/~~LF)} \end{tabular} &  \begin{tabular}[c]{@{}c@{}}0.445 \\(0.410~~/~~0.035)\end{tabular} &
\begin{tabular}[c]{@{}c@{}}0.345 \\(0.276~~/~~0.069)\end{tabular} &
 \begin{tabular}[c]{@{}c@{}}0.309 \\(0.132~~/~~0.176)\end{tabular}              & 0.346             & 0.339              & 0.338              \\ 
\hline
\textbf{PSNR (dB)}        & 
32.38                  & 32.35                   & 31.35                   & 32.08             & 31.86              & 28.92              \\ 
\hline
\textbf{MS-SSIM (dB)}     & 
15.26                  & 15.25                   & 15.08                   & 14.99             & 14.34              & 12.37              \\ 
\hline
\textbf{FLOPs (G)}  & 
16.57 & 13.69 & 11.04 & 12.98 & 11.51 & 8.84 \\ 
\hline
\end{tabular}
\label{table:ablation_study}
\end{table*}

One visual example from the Kodak image set is given in Figure \ref{fig:kodak_quan1} in which our results are qualitatively compared with JPEG2000 and BPG (4:4:4 chroma format) at 0.1bpp. As seen in the example, our method provides the highest visual quality compared to the others. JPEG2000 has poor performance due to the ringing artifacts. The BPG result is smoother compared to JPEG2000, but the details and fine structures are not preserved in many areas, for example, in the patterns on the shirt and the colors around the eye.

\subsection{Ablation Study}

In order to evaluate the performance of different components of the proposed framework, ablation studies were performed, which are reported in Table \ref{table:ablation_study}. The results are the average over the Kodak image set. All the models reported in this ablation study have been optimized for MSE distortion metric (for one single bit-rate). However, the results were evaluated with both PSNR and MS-SSIM metric.
\begin{itemize}
\item
\textbf{Ratio of HF and LF}: in order to study varying choices of the ratio of channels allocated to the LF feature representations, we evaluated our model with three different ratios $\alpha \in \{0.25,0.5,0.75\}$. As summarized in Table \ref{table:ablation_study}, compressing 50\% of the LF part to half the resolution (i.e., $\alpha=0.5$) results in the best R-D performance in both PSNR and MS-SSIM at 0.345bpp (where the contributions of HF and LF latents are 0.276bpp and 0.069bpp). 
As the ratio decreases to $\alpha=0.25$, less compression with a higher bit rate of 0.445bpp (0.410bpp for HF and 0.035 for LF) is obtained, while no significant gain in the reconstruction quality is achieved. Although increasing the ratio to 75\% provides a better compression with 0.309bpp (high: 0.132bpp, low: 0.176bpp), it significantly results in a lower PSNR. As indicated by the number of floating point operations per second (FLOPs) in the table, larger ratio results in a faster model since less operations are required for calculating the LF maps with half spatial resolution.
\item
\textbf{Position of activation layers}: in this scenario (denoted by ActOut), we cancel the internal activations (i.e., GDN/IGDN) employed in our proposed GoConv and GoTConv. Instead, as in the original octave convolution \cite{chen2019drop}, we apply GDN to the output HF and LF maps in GoConv, and IGDN before the input HF and LF maps for GoTConv. This experiment is denoted by ActOut in Table \ref{table:ablation_study}. As the comparison results indicate, the proposed architecture with internal activations ($\boldsymbol{\alpha}$ = 0.5) provides a better performance (with $\approx$0.27dB higher PNSR) since all internal feature maps corresponding to the inter- and intra-communications are benefited from the activation function.
\item
\textbf{Octave only for core auto-encoder}: as described in Section \ref{sec:Multi-Frequency Entropy Model}, the proposed multi-frequency entropy model utilizes GoConv and GoTConv units for both latents and hyper latents. In order to evaluate the effectiveness of multi-frequency modelling of hyper latents, we also report the results in which GoConv and GoTConv are only used for the core auto-encoder latents (denoted by CoreOct in Table \ref{table:ablation_study}). To deal with the HF and LF latents resulted from the multi-frequency core auto-encoder, we used two separate networks (similar to \cite{minnen2018joint} with Vanilla convolutions) for each of the hyper encoder, and hyper decoder. As summarized in the table, a PSNR gain of $\approx$0.49dB is achieved when both core and hyper auto-encoders benefit from the proposed multi-frequency model.

\begin{figure}[htb!]
\centering
  \centerline{
  \includegraphics[width=0.95\linewidth]{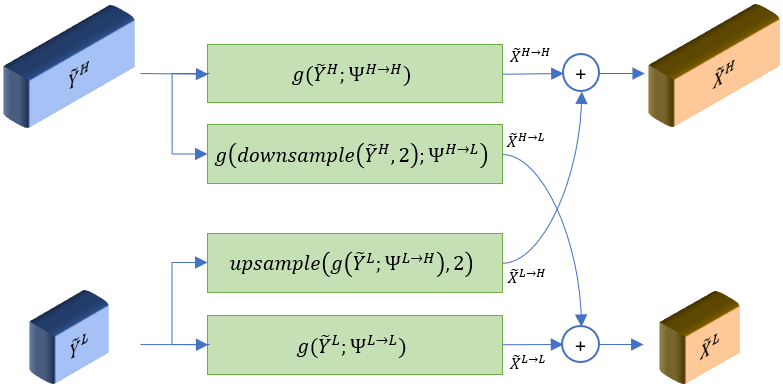}}
\caption{The architecture of the original octave transposed-convolution. $\Tilde{Y}^H$ and $\Tilde{Y}^L$: input HF and LF feature maps; $g$: regular vanilla transposed-convolution; $upsample$: fixed up-sampling operation (i.e., bilinear); $downsample$: fixed down-sampling operation (i.e., maxpooling); $\Tilde{X}^{H\rightarrow H}$ and $\Tilde{X}^{L\rightarrow L}$: intra-frequency updates; $\Tilde{X}^{H\rightarrow L}$ and $\Tilde{X}^{L\rightarrow H}$: inter-frequency communications; $\Psi^{H\rightarrow H}$ and $\Psi^{L\rightarrow L}$: intra-frequency transposed-convolution kernels; $\Psi^{H\rightarrow L}$ and $\Psi^{L\rightarrow H}$: inter-frequency transposed-convolution kernels; $\Tilde{X}^H$ and $\Tilde{X}^L$: output HF and LF feature maps.}
\label{fig:octtconv}
\end{figure}

\item
\textbf{Original octave convolutions}: in this experiment, the performance of the proposed GoConv and GoTConv architectures compared with the original octave convolutions (denoted by OrgOct in Table \ref{table:ablation_study}) is analyzed. We replace all GoConv layers in the proposed framework (Figure \ref{fig:framework}) by original octave convolutions (Figure \ref{fig:octconv}). For the octave transposed-convolution used in the core and hyper decoders, we reverse the octave convolution operation formulated as follows:
\begin{equation}
\begin{split}
       \Tilde{X}^{H} &= g(\Tilde{Y}^H;\Psi^{H\rightarrow H})+upsample(g(\Tilde{Y}^L;\Psi^{L\rightarrow H}),2), \\
       \Tilde{X}^{L} &= g(\Tilde{Y}^L;\Psi^{L\rightarrow L})+g(downsample(\Tilde{Y}^H,2);\Psi^{H\rightarrow L}),
\end{split}
\end{equation}
where $\{\Tilde{Y}^{H},\Tilde{Y}^{L}\}$ and $\{\Tilde{X}^{H},\Tilde{X}^{L}\}$ are the input and output feature maps, and $g$ is vanilla transposed-convolution. The architecture of the original octave transposed-convolution is illustrated in Figure \ref{fig:octtconv}. Similar to the octave convolution defined in \cite{chen2019drop}, average pooling and nearest interpolation are respectively used for down- and up-sampling operations. 

As reported in Table \ref{table:ablation_study}, OrgOct provides a significantly lower performance than the architecture with the proposed GoConv/GoTConv, which is due to the fixed sub-sampling operations incorporated for its inter-frequency components. The PSNR and MS-SSIM of the proposed architecture are respectively $\approx$3.43dB and $\approx$2.88dB higher than Org-Conv at the same bit rate. Compared to the other models, OrgOct has the lowest complexity with respect to FLOPs. Note that the ratio $\alpha=0.5$ was used for the ActOut, CoreOct, and OrgOct models.
\end{itemize}


\begin{figure*}
\centering
\begin{subfigure}[b]{.24\textwidth}
 \centering
  \centerline{\includegraphics[width=\textwidth]{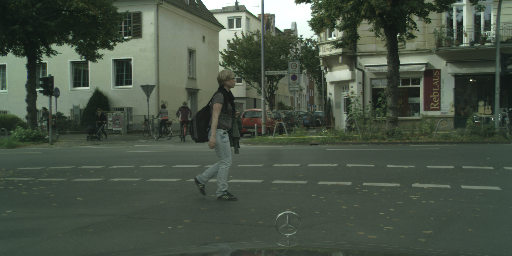}}
  \subcaption{\scriptsize Input Image}
\end{subfigure}
\begin{subfigure}[b]{.24\textwidth}
 \centering
  \centerline{\includegraphics[width=\textwidth]{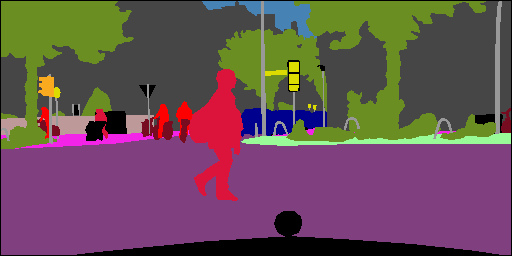}}
  \subcaption{\scriptsize Ground Truth}
\end{subfigure}
\begin{subfigure}[b]{.24\textwidth}
 \centering
  \centerline{\includegraphics[width=\textwidth]{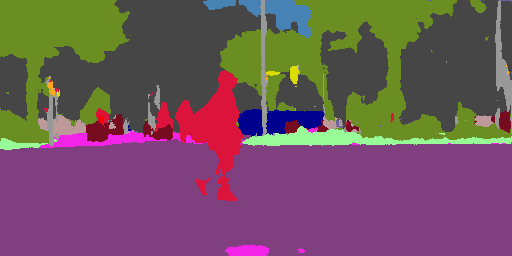}}
  \subcaption{\scriptsize GoConv-UNet (Acc: 0.86, mIoU: 0.37)}
\end{subfigure}
\begin{subfigure}[b]{.24\textwidth}
 \centering
  \centerline{\includegraphics[width=\textwidth]{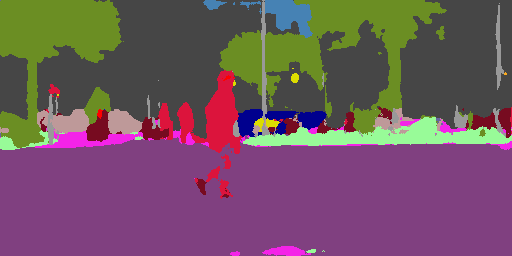}}
  \subcaption{\scriptsize UNet (Acc: 0.84, mIoU: 0.31)}
\end{subfigure}
\vskip 5pt
\begin{subfigure}[b]{.24\textwidth}
 \centering
  \centerline{\includegraphics[width=\textwidth]{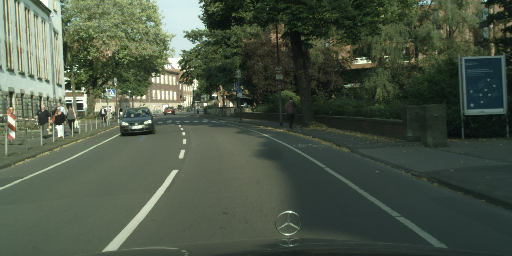}}
  \subcaption{\scriptsize Input Image}
\end{subfigure}
\begin{subfigure}[b]{.24\textwidth}
 \centering
  \centerline{\includegraphics[width=\textwidth]{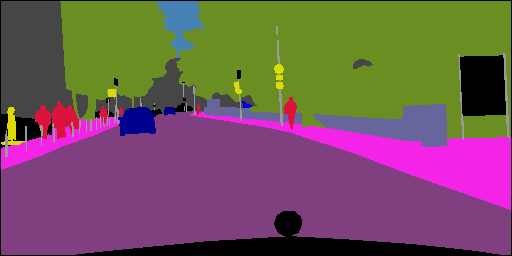}}
  \subcaption{\scriptsize Ground Truth}
\end{subfigure}
\begin{subfigure}[b]{.24\textwidth}
 \centering
  \centerline{\includegraphics[width=\textwidth]{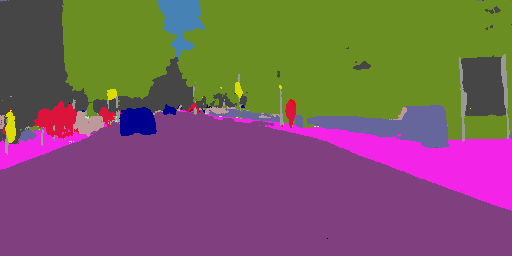}}
  \subcaption{\scriptsize GoConv-UNet (Acc: 0.89, mIoU: 0.38)}
\end{subfigure}
\begin{subfigure}[b]{.24\textwidth}
 \centering
  \centerline{\includegraphics[width=\textwidth]{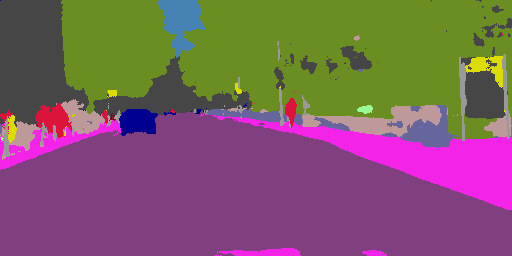}}
  \subcaption{\scriptsize UNet (Acc: 0.86, mIoU: 0.34)}
\end{subfigure}
\caption{Visual examples on Cityscapes. \textbf{GoConv-UNet}: GoConv/GoTConv-based multi-frequency UNet with $\alpha=0.125$; \textbf{UNet}: original UNet architecture; \textbf{Acc}: pixel accuracy; \textbf{mIoU}: mean IoU.}
\label{fig:semantic_visual2}
\end{figure*}

As explained in Section \ref{sec:Proposed Generalized octave Convolution}, since the proposed GoConv and GoTConv are designed as generic, plug-and-play units, they can be used in any auto-encoder-based architecture. In the following sections, the usage of GoConv and GoTConv in image semantic segmetnation and image denoising problems are analyzed.

\section{Multi-Frequency Semantic Segmentation}
\label{ssec:Multi-Frequency Semantic Segmentation}

In this experiment, we evaluate the proposed GoConv/GoTConv units on image semantic segmentation. The popular UNet model \cite{ronneberger2015u} is used as the baseline in this experiment. UNet is a CNN-based architecture originally developed for segmenting biomedical images, but also considered as a baseline for image semantic segmentation. UNet has an auto-encoder scheme composed of two paths: the contraction path (i.e., the encoder) and the expanding path (i.e., the decoder). The encoder captures the context in the image, while the decoder is used to enable precise localization. The architecture of UNet auto-encoder is summarized in Table \ref{table:unet}. All the vanilla convolution layers used in the encoder and decoder are  followed by batch normalization and ReLU layers.

In this study, we build a multi-frequency UNet model, denoted by GoConv-UNet, by replacing all the UNet convolution layers with GoConv and all transposed-convolutions with GoTConv. The other properties such as the number of layers, filters, and strides are kept the same as the original UNet. In order to compare the performance of the original octave convolutions with our proposed GoConv/GoTConv, we build another model, denoted by OrgOct-UNet, in which the original octave convolution units (shown in Figures \ref{fig:octconv} and \ref{fig:octtconv}) are employed. 

All the experiments in this section are performed on Cityscapes dataset \cite{Cordts2016Cityscapes}, which contains 2974 training images with 19 semantic labels. All models were trained with cross-entropy loss function for 100 epochs. The models were then evaluated on Cityscapes validation set (including 500 images) based on pixel accuracy, mean intersection-over-union (IoU), number of parameters, and number of FLOPs. The mean IoU (mIou) is calculated by averaging over the IoU values of all semantic classes.

Table \ref{table:semseg_results} presents the comparison results of the original UNet and the proposed multi-frequency OrgOct-UNet and GoConv-UNet models with different $\alpha$'s. All the reported values are the average over the Cityscapes validation set. As given in the table, the GoConv-UNet model with $\alpha=0.125$ achieves the highest accuracy and mIoU. Compared to the original UNet model, GoConv-UNet ($\alpha=0.125$) requires $\approx$0.7M less parameters, and achieves 25\% FLOPs reduction, which clearly shows the benefit of the proposed GoConv and GoTConv units in multi-frequency semantic segmentation. 

\begin{table}
\centering
\caption{UNet architecture (\textbf{Conv}: vanilla convolution; \textbf{T-Conv}: vanilla transposed-convolution).}
\begin{tabular}{|c|c|} 
\hline
\textbf{Encoder}    & \textbf{Decoder}       \\ 
\hline
\centering
Conv (3*3, 64, s1)  & T-Conv (3*3, 512, s2)  \\
Maxpool (2*2, s2)  & Conv (3*3, 512, s1)  \\
Conv (3*3, 128, s1)  & T-Conv (3*3, 256, s1)  \\
Maxpool (2*2, s2)  & Conv (3*3, 256, s1)  \\
Conv (3*3, 256, s1)  & T-Conv (3*3, 128, s1)   \\
Maxpool (2*2, s2)  & Conv (3*3, 128, s1)  \\
Conv (3*3, 512, s1)  & T-Conv (3*3, 64, s2)   \\
Maxpool (2*2, s2)  & Conv (3*3, 64, s1)  \\
Conv (3*3, 1024, s1)  & Conv (3*3, 19, s1)   \\
\hline
\end{tabular}
\label{table:unet}
\end{table}

Although larger $\alpha$'s can result in lower FLOPs, it has a negative impact on the model performance. OrgOct-UNet has the lowest number of parameters and FLOPs compared to the other models. However, it achieves the lowest performance with no improvement over UNet, which is basically due to the fixed sub-sampling operations used in the original octave convolutions.

The visual comparison of UNet and GoConv-UNet is given in Figure \ref{fig:semantic_visual2}. As shown in the results, higher quality semantic segmentations  are obtained with GoConv-UNet. For example, the pedestrian with backpack in the top example and the trees, the board, and the wall on the right side of the bottom example are more accurately segmented with GoConv-UNet.



\begin{table*}
\centering
\caption{Comparison results for image semantic segmentation on Cityscapes database. 
\textbf{UNet}: original UNet architecture; \textbf{GoConv-UNet}: multi-frequency UNet with GoConv/GoTConv;  \textbf{OrgOct-UNet}: multi-frequency UNet with original octave units; 
\textbf{Ratio ($\alpha$)}: the LF ratio used in octave convolutions; \textbf{mIoU}: mean intersection-over-union; \textbf{Params (M)}: the number of model parameters (in million)}
\label{table:semseg_results}
\begin{tabular}{|l|c|c|l|l|c|l|} 
\cline{2-7}
\multicolumn{1}{c|}{} & \textbf{UNet}               & \multicolumn{3}{c|}{\textbf{GoConv-UNet} }           & \multicolumn{2}{c|}{\textbf{OrgOct-UNet} }  \\ 
\hline
\textbf{Ratio ($\alpha$)}  & 0.0 & 0.5                        & 0.25  & 0.125 & 0.5                        & 0.125           \\ 
\hline
\textbf{Accuracy} & 0.787 & 0.781 & 0.789 & \textbf{0.791}  & 0.776 & 0.784 \\ 
\hline
\textbf{mIoU}                  & 0.402                       & 0.381                      & 0.401 & \textbf{0.414}  & 0.379                      & 0.399           \\ 
\hline
\textbf{Params (M)}  & 34.54 & 32.96 & 33.35 & 33.84 & 32.85                      & 33.73           \\ 
\hline
\textbf{FLOPs (G)}                & \multicolumn{1}{l|}{103.44} & \multicolumn{1}{l|}{46.77} & 70.60 & 85.80           & \multicolumn{1}{l|}{45.64} & 85.46           \\
\hline
\end{tabular}
\end{table*}

\section{Multi-Frequency Image Denoising}
\label{ssec:Multi-Frequency Image Denoising}

In this experiment, we build a simple convolutional auto-encoder and use it for image denoising problem \cite{liu2018multi,tian2019deep}. In this problem, we try to denoise images corrupted by white Gaussian noise, which is the common result of many acquisition channels. The architecture of the auto-encoder used in this experiment is summarized in Table \ref{table:image denoising network} where the encoder and decoder are respectively composed of a sequence of vanilla convolutions and transposed-convolutions each followed by batch normalization and ReLU. 

We performed our experiments on MNIST and CIFAR10 datasets. After 100 epochs of training, an average PSNR of 23.19dB and 23.29dB for MNIST and CIFAR10 test sets are respectively achieved. In order to analyze the performance of GoConv and GoTConv in this experiment, we replaced all the vanilla convolution layers with GoConv and all transposed-convolutions with GoTConv. The other properties of the encoder and decoder networks (e.g., numebr of layers, filters, and strides) are the same as the baseline in Table \ref{table:image denoising network}. We set $\alpha=0.125$ and trained the model for 100 epochs. 

For MNIST dataset, the multi-frequency auto-encoder achieved an average PSNR of 23.20dB (almost the same as the baseline with vanilla convolutions). However, for CIFAR10, we achieved an average PSNR of 23.54dB, which is 0.25dB higher than the baseline, due to the effective communication between HF and LF. In addition, the proposed multi-frequency auto-encoder has less number of parameters and FLOPs than the baseline model, which indicates the benefit of octave convolutions on parameters and operations reduction with no performance loss. The comparison results are presented in Table \ref{table:results}.

\begin{figure}
\centering
\begin{subfigure}[b]{.48\textwidth}
 \centering
  \centerline{\includegraphics[width=\textwidth]{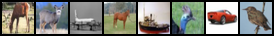}}
  \caption{Input Images}
\end{subfigure}
\begin{subfigure}[b]{.48\textwidth}
 \centering
  \centerline{\includegraphics[width=\textwidth]{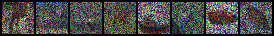}}
  \caption{Input Noisy Images}
\end{subfigure}
\begin{subfigure}[b]{.48\textwidth}
 \centering
  \centerline{\includegraphics[width=\textwidth]{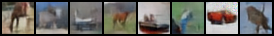}}
  \caption{Baseline Denoised Results}
\end{subfigure}
\begin{subfigure}[b]{.48\textwidth}
 \centering
  \centerline{\includegraphics[width=\textwidth]{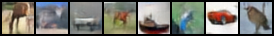}}
  \caption{Multi-frequency Denoised Results}
\end{subfigure}
\caption{Sample image denoising results from CIFAR10.}
\label{fig:visual examples}
\end{figure}

In Figure \ref{fig:visual examples}, 8 visual examples from CIFAR10 test set are given. Compared to the baseline model with vanilla convolutions and transposed-convolutions, the multi-frequency model with the proposed GoConv/GoTConv results in a higher visual quality in the denoised images (e.g., the red car in the second column from right).

\begin{table}
\centering
\caption{Baseline convolutional auto-encoder for image denoising (\textbf{Conv}: vanilla convolution; \textbf{T-Conv}: vanilla transposed-convolution).}
\begin{tabular}{|c|c|} 
\hline
\textbf{Encoder}    & \textbf{Decoder}       \\ 
\hline
\centering
Conv (3*3, 32, s1)  & T-Conv (3*3, 128, s2)  \\
Conv (3*3, 32, s1)  & T-Conv (3*3, 128, s1)  \\
Conv (3*3, 64, s1)  & T-Conv (3*3, 64, s1)   \\
Conv (3*3, 64, s2)  & T-Conv (3*3, 64, s2)   \\
Conv (3*3, 128, s1) & T-Conv (3*3, 32, s1)   \\
Conv (3*3, 128, s1) & T-Conv (3*3, 32, s1)   \\
Conv (3*3, 256, s2) & T-Conv (3*3, 3, s1)    \\
\hline
\end{tabular}
\label{table:image denoising network}
\end{table}

\begin{table}
\centering
\caption{Comparison results of the baseline and multi-frequency auto-encoders for image denoising on MNIST and CIFAR10 test sets.}
\label{table:results}
\begin{tabular}{|l|c|c|c|c|} 
\cline{2-5}
\multicolumn{1}{c|}{} & \multicolumn{2}{c|}{\textbf{Baseline} } & \multicolumn{2}{c|}{\textbf{Multi-frequency} }  \\ 
\cline{2-5}
\multicolumn{1}{c|}{} & MNIST & CIFAR10                         & MNIST           & CIFAR10                       \\ 
\hline
\textbf{PSNR (dB)}   & 23.19 & 23.29                           & \textbf{23.20}  & \textbf{23.54}                \\ 
\hline
\textbf{Params (M)}            & 1.16  & 1.17                            & \textbf{1.10}   & \textbf{1.14}                 \\ 
\hline
\textbf{FLOPS (G)} & 0.20  & 0.23                            & \textbf{0.17}   & \textbf{0.20}                 \\ 
\hline
\end{tabular}
\end{table}

\begin{figure}
\centering
\begin{subfigure}[b]{.48\textwidth}
 \centering
  \centerline{\includegraphics[width=\textwidth]{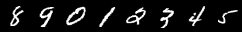}}
  \caption{Input Images}
\end{subfigure}
\begin{subfigure}[b]{.48\textwidth}
 \centering
  \centerline{\includegraphics[width=\textwidth]{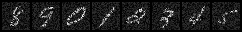}}
  \caption{Input Noisy Images}
\end{subfigure}
\begin{subfigure}[b]{.48\textwidth}
 \centering
  \centerline{\includegraphics[width=\textwidth]{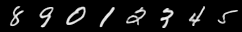}}
  \caption{Baseline Denoised Results}
\end{subfigure}
\begin{subfigure}[b]{.48\textwidth}
 \centering
  \centerline{\includegraphics[width=\textwidth]{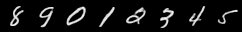}}
  \caption{Multi-frequency Denoised Results}
\end{subfigure}
\caption{Sample image denoising results from MNIST test set.}
\label{fig:mnist visual examples}
\end{figure}

\section{Conclusion}
\label{concolusion}

In this paper, we propose a new learned multi-frequency image compression and entropy model with octave convolutions in which the latents are factorized into HF and LF components, and the LF is stored at lower resolution to reduce the spatial redundancy. To preserve the spatial structure of the input, novel generalized octave convolution and transposed-convolution architectures denoted by GoConv/GoTConv are introduced. Our experiments show that the proposed method significantly improves the R-D performance and achieves the new state-of-the-art learned image compression, which even outperforms VTM (4:2:0) in PSNR. Further improvements can be achieved by  multi-frequency factorization of latents into a sequence of high to low frequencies as in wavelet transform.

Our method bridges the wavelet transform and deep learning-based image compression, and allows many techniques in the wavelet transform research to be applied to learned image compression. This will lead to many other research topics in the future. We also show the benefit of the proposed GoConv/GoTConv in other CNN-based computer vision applications, particularly auto-encoder-based schemes such as image denoising and semantic segmentation. 


%

\section*{Acknowledgements}
This work is supported by the Natural Sciences and Engineering Research Council (NSERC) of Canada under grant RGPIN-2015-06522.


\bibliography{refs}
\bibliographystyle{IEEEtran}

\end{document}